\documentclass[conference, a4paper]{IEEEtran}
\IEEEoverridecommandlockouts

\usepackage{cite}

\ifCLASSINFOpdf
  \usepackage[pdftex]{graphicx}
  \DeclareGraphicsExtensions{.pdf,.jpeg,.jpg,.png}
  \usepackage[protrusion=true,expansion=true]{microtype}  
\usepackage[pdftex,colorlinks,linkcolor=black,citecolor=black,urlcolor=black,hyperfootnotes=false]{hyperref}

\else
  \usepackage[dvips]{graphicx}
  \DeclareGraphicsExtensions{.eps,.jpeg,.jpg,.png}
  \usepackage{microtype}
\fi
\usepackage{amsmath}
\usepackage{amssymb}
\usepackage{algorithm}
\usepackage[noend]{algpseudocode}

\usepackage{array}

\usepackage{todonotes}
\usepackage{multirow}
\usepackage{gensymb} 

\newcommand{\integral}[4]{
\int_{#3}^{#4}  #1 \, \mathrm{d}#2 }

\newcommand{\e}[1]{ \mathrm{e}^{#1}}

\renewcommand{\log}[1]{ \mathrm{log}\left( #1 \right)}

\newcommand{\Ton}[0]{T_{\mathrm{on}}}
\newcommand{\Toff}[0]{T_{\mathrm{off}}}
\newcommand{\Tmax}[0]{T_{\mathrm{max}}}
\newcommand{\Tmin}[0]{T_{\mathrm{min}}}
\newcommand{\Tlow}[0]{T_{\mathrm{low}}}
\newcommand{\Thigh}[0]{T_{\mathrm{high}}}

\begin{document}

%
\title{Low-complexity control algorithm for decentralised demand response using thermostatic loads}

\author{\IEEEauthorblockN{Simon Tindemans}
\IEEEauthorblockA{Department of Electrical Sustainable Energy\\
Delft University of Technology\\
Delft, The Netherlands\\
s.h.tindemans@tudelft.nl}
\and
\IEEEauthorblockN{Goran Strbac}
\IEEEauthorblockA{Department of Electrical and Electronic Engineering\\
Imperial College London\\
London, United Kingdom\\
g.strbac@imperial.ac.uk}
\thanks{This research was supported by EDF Energy R\&D UK Centre as part of the STAMINA project.}}


%


\maketitle

\begin{abstract}
Thermostatically controlled loads such as refrigerators are exceptionally suitable as a flexible demand resource. This paper derives a decentralised load control algorithm for refrigerators. It is adapted from an existing continuous time control approach, with the aim to achieve low computational complexity and an ability to handle discrete time steps of variable length -- desirable features for embedding in appliances and high-throughput simulations.  Simulation results of large populations of heterogeneous appliances illustrate the accurate aggregate control of power consumption and high computational efficiency. 
\end{abstract}

\renewcommand\IEEEkeywordsname{Keywords}

\begin{IEEEkeywords}
thermostatically controlled loads, demand response, decentralized control, aggregate control
\end{IEEEkeywords}

%
\IEEEpeerreviewmaketitle

\section{Introduction}

The physical characteristics of refrigerators and other thermostatically controlled loads  (TCLs) make them exceptionally suitable as a low-cost provider of flexibility to the grid: their power consumption can be shifted by tens of minutes without noticeable effects on cooling performance. This flexibility can then be used for the provision of response and reserve services, to reduce extreme load levels and to alleviate ramping constraints \cite{Callaway2011}. Given the large number of devices involved, large populations can be effectively controlled in a decentralised manner using randomised control schemes, as proposed e.g. in \cite{Angeli2012, Meyn2015, Totu2017}. Practical implementation must also consider constraints on implementation, computation and communication, as discussed e.g. in
\cite{Iacovella2016, PoncedeLeonBarido2017}.

A robust decentralised control scheme for heterogeneous TCLs was introduced in \cite{Tindemans2015}, and extended in \cite{Trovato2015} to allow for scenarios in which TCLs collectively absorb energy from the grid. The control strategy has the desirable feature that it requires only one-way broadcast information, yet achieves tracking of a reference signal that is exact in expectation (i.e. exact for large numbers) without violating temperature limits in individual devices. However, its continuous time formulation in integral form was not conducive to implementation on embedded controllers or for rapid simulation of many devices at once. 
 
This paper addresses that shortcoming by deriving a discrete-time control algorithm that implements the control strategy that was presented in \cite{Tindemans2015, Trovato2015}. The algorithm is particularly suitable for implementation on devices with computational constraints. Specifically, it avoids numerical integration and uses greedy time steps of variable size, so that real-time performance requirements are relaxed. First, a discretisation procedure is described, for the case of a piecewise-constant control signal, and the controller is reformulated in natural coordinates. Then, expressions are derived for each of the on/off-switching processes governing the behaviour of the device. Finally, an explicit control algorithm is provided, and its efficacy is illustrated with Python-based simulations for a heterogeneous population of refrigerators.

\section{Preliminaries}

\subsection{Appliance model}

Throughout this paper, we consider the first order TCL model, expressed by the following differential equation for the temperature $T^a$ of the compartment of appliance $a$:
\begin{equation}
\frac{\mathrm{d} T^a(t)}{\mathrm{d}t} = - \alpha^a \left[ T^a(t) - T^a_{\mathrm{off}} + c^a(t) \cdot \left(  T^a_{\mathrm{off}} - T^a_{\mathrm{on}} \right)  \right]
\end{equation}
For ease of exposition, we shall refer to refrigerating applicances throughout, although the same model and control strategy can be used for other TCLs, e.g. space heaters. Here, $c^a(t) \in \{0,1\}$ is the state of the compressor, and $T^a_{\mathrm{off}}$ and $T^a_{\mathrm{on}}$ are the asymptotic temperatures in the \texttt{off} and \texttt{on} states, respectively. The power consumption $P^a(t)$ of the appliance is assumed to be dominated by the compressor power consumption $P^a_{\textrm{on}}$, so that
\begin{equation}
P^a(t) = P^a_{\textrm{on}} c^a(t).
\end{equation}

In the steady state (no control actions), the appliance is subject to a hysteresis controller that switches to the \texttt{on} state whenever an upper temperature bound $\Tmax^a$ is reached, and to the \texttt{off} state when $\Tmin^a$ is reached. This results in a periodic cycling of the power consumption, with an average power level $P^a_0$. Let us consider that each appliance $a$ has a model $\mathcal{M}^a =\{ \alpha^a, P^a_{\textrm{on}}, , \Toff^a, \Ton^a, \Tmin^a, \Tmax^a\}$ that is known to us, but an internal state $\mathcal{S}^a = \{T^a(t), c^a(t)\}$ that is not. We assume the latter is distributed according to a steady state distribution. Then, the power consumption is in effect a random process, with the expectation (at each time $t$)
\begin{equation} \label{eq:expectationsingle}
E_{\mathcal{S}^a}[P^a(t)] = P^a_0.
\end{equation}

\subsection{Aggregate power modulation}

The objective of the TCL demand response controller is to control the aggregate power consumption 
\begin{equation} \label{eq:sumpower}
P(t) = \sum_{a \in \mathcal{A}} P^a(t)
\end{equation}
of a collection of appliances $\mathcal{A}$. In this paper we consider the control approach introduced in \cite{Tindemans2015}, which modulates the power consumption using a broadcast reference signal $\Pi(t)$ in the following way. Each device individually adapts its power consumption to $\Pi(t)$ in order to satisfy 
\begin{equation} \label{eq:expectationpi}
E_{\mathcal{S}^a}[P(t)] = \Pi(t) P^a_0.
\end{equation}
Clearly, $\Pi(t)=1$ represents the steady state \eqref{eq:expectationsingle} and changes in $\Pi(t)$ are immediately reflected in the expected power consumption. Moreover, the controller maintains independence between appliances (conditional on the control signal $\Pi(\cdot)$) so that the central limit theorem can be applied to the total power consumption \eqref{eq:sumpower}, resulting in
\begin{equation}
P(t) = \Pi(t) \sum_{a \in \mathcal{A}} P^a_0 + O(\sqrt{|\mathcal{A}|}),
\end{equation}
where the last term is a random process that decreases in relative importance to the first term as the set of appliances increases. We note that this is the case even for heterogeneous appliances. 

The ability to closely track a reference signal was first demonstrated in \cite{Tindemans2015}. In \cite{Trovato2015}, the control signal was generated using a mixture of off-line scheduling and real-time control, and in \cite{Tindemans2015a} various frequency-sensitive controllers to locally compute $\Pi(t)$ (e.g. a simple droop controller) were implemented.

\subsection{Distribution-referred control} \label{sec:drcontrol}

The control approach introduced in \cite{Tindemans2015} and extended in \cite{Trovato2015} is\emph{distribution-referred} approach, because it implements control via the probability distribution of temperatures for appliances with a known model $\mathcal{M}^a$ and unknown state $\mathcal{S}^a$. It defines a family of alternative distributions $f_z(T;\mathcal{M}^a)$ that varies continuously in the parameter $z$, containing as a special case the steady state temperature distribution \cite[Eqs.~(31)-(32)]{Tindemans2015}
\begin{equation}
f_0(T; \mathcal{M}^a) = \frac{k^a}{(\Toff^a - T)(T-\Ton^a)}
\end{equation}
with 
\begin{equation} \label{eq:c-constant}
k^a = \frac{\Toff^a-\Ton^a}{\log{ \frac{(\Tmax^a-\Ton^a)
   (\Tmin^a-\Toff^a)}{(\Tmin^a-\Ton^a)(\Tmax^a-\Toff^a) }}}
\end{equation}

The controller consists of two major elements, which are evaluated by each appliance in order:
\begin{enumerate}
\item Choose the distribution parameter $z(t)$ such that the power consumption tracks the reference signal $\Pi(t)$ according to \eqref{eq:expectationpi}. Determine the collective device switching actions required to keep device temperatures aligned with $f_{z(t)}(T)$, and identify the temperature limits $\Tmin(t) \ge \Tmin$ and $\Tmax(t) \le \Tmax$.
\item Based on the \emph{actual} appliance state $\mathcal{S}^a$, compute stochastic control actions, in the form of \texttt{on}/\texttt{off} switching. Switching events can be initiated in three distinct ways:
\begin{itemize}
\item Deterministic switching when temperature limits $\Tmin(t)$ or $\Tmax(t)$ are exceeded.
\item Continuous-time stochastic switching at intermediate temperatures in order to shape the temperature distribution.
\item Instantaneous stochastic switching on discrete changes of power setpoints, or when the controller switches between energy-provision and energy-absorption modes \cite{Trovato2015}.
\end{itemize}
\end{enumerate}
The temperature distribution and appliance switching phases for the discrete time control strategy are addressed in Sections \ref{sec:distribution} and \ref{sec:switching}, respectively. We will henceforth drop the appliance superscript $a$, because the control steps are executed locally within each appliance (or independently for each appliance in a simulation).  Note that this implies a single model $\mathcal{M}$ is used in the derivations, but the results remain valid for portfolios of heterogeneous devices (each with their own model).

\subsection{Discretisation procedure}

Moving from a continuous time formulation to a discrete time formulation, we consider a partitioning of the timeline by the ordered sequence of times $\{t_i\}$, indexed by the integer $i$, at which the controller is invoked. They define time intervals $(t_{i+1}, t_i]$ with durations $\Delta t_i =t_i - t_{i-1}$. Note that the duration $\Delta t_i$ refers to the interval \emph{prior} to $t_i$, and the intervals may have variable size. The reference signal $\Pi(t)$ is assumed to be piecewise constant, defined by 
\begin{equation}
\Pi(t) = \Pi_i, \quad \text{for } t\in (t_{i-1}, t_i].
\end{equation}
The controller thus receives at $t_i$ a new reference power level $\Pi_{i+1}$ that must be applied for the upcoming interval $(t_i,t_{i+1}]$. 

Although the discontinuous changes of reference power will trigger switching events at $t_i$, the other switching events may occur at any time $t$. In the discretised approximation of the continuous time controller, they will be synchronised with the control execution times $t_i$ as follows. It is assumed that switching is immediate (at $t_i$).
\begin{itemize}
\item A violation of the temperature limits will trigger corrective switching as soon as it is detected.
\item Switching due to toggling between energy-provision and energy-absorption modes is implemented as soon as a change in regime is detected.
\item Continuous time stochastic switching is implemented by approximating the integrated switching rate (i.e. the switching probability) over $\Delta t_i$ using the trapezoidal method, and executing any switching events at $t_i$ (the end of the interval). 
\end{itemize}

The algorithm is thus implemented in a `backward' fashion, meaning that at time $t_i$, the algorithm implements switching actions resulting from reference changes at $t_i$, and those accumulated over the \emph{preceding} interval $(t_{i-1}, t_i]$. The advantage of this approach is that the interval $\Delta t_i$ can be chosen opportunistically: the controller does not need to know in advance when the next time step will take place. This is convenient, for example when computational limitations cause a delay in intended invocation time, or when the time step adapts to sudden changes in grid frequency. It should be pointed out that this `backward' integration does not delay the response to changes in reference power, which is implemented immediately at $t_i$. 

\section{Distribution behaviour} \label{sec:distribution}

This section focuses on the first part of the controller. It computes the desired evolution of the probability distribution of temperatures of fridges with model $\mathcal{M}$, when tracking a piecewise constant reference $\Pi_i$. The derivation is initially performed in continuous time. The results are subsequently expressed in natural coordinates and restated in a form that is suitable for discrete-time evaluation. 

\subsection{Aggregate physics}

The average temperature of a TCL population is affected by the desired power consumption $\Pi(t)$ according to \cite[Eq.~(26)]{Tindemans2015}. With the convention that $t_{-1}=-\infty$ and $\Pi_0 = 1$ (assuming an initial steady state), it follows that
\begin{align}
\overline{T}(t_i) 
&= \Toff - \alpha (\Toff - \overline{T}_0) \sum_{j=0}^{i} \Pi_j \integral{ \e{-\alpha (t_i-t')}}{t'}{t_{j-1}}{t_j}, \label{eq:avgtempdefinition}
\end{align}
where the steady state average temperature $\overline{T}_0$ is computed using \cite[Eqs.~(23) and (32)-(33)]{Tindemans2015} as
\begin{equation}
\overline{T}_0 = \Toff - k \times \log{\frac{\Tmax-\Ton}{\Tmin - \Ton}} \label{eq:steadyStateTemperature}
\end{equation}
with $k$ defined in \eqref{eq:c-constant}. We define the dimensionless variable
\begin{equation}
z(t) =\frac{\overline{T}_0 - \overline{T}(t)}{\Toff - \overline{T}_0}, \label{eq:zdefinition}
\end{equation}
to parameterise the distributions $f_{z(t)}(T)$, and simplify the notation in what follows. Note that it is related to the $\sigma$ variable used in \cite{Trovato2015} as $z=\sigma-1$.

\subsection{Controller modes}

The algorithm in \cite{Tindemans2015} implicitly generates the family of temperature distributions $f_{z(t)}(T)$ by the net heating rate $v(T,t)$, which is determined by averaging over devices in the \texttt{off} (heating) and \texttt{on} (cooling) states at time $t$ and temperature $T$. The heating rate is controlled by a parameter $\beta(t)$ through $v(T,t) =\alpha \beta(t) (T - \Tmax)$. The effect of this heating rate profile is a temperature distribution that contracts to the pivot temperature $\Tmax$ in order to provide energy to the grid - and reverses this process to recover the energy supplied. In \cite{Trovato2015} it was coupled to a `mirrored' controller that is capable of absorbing energy from the grid by contracting to the pivot temperature $\Tmin$. Switching between the two controller modes takes place whenever $\overline{T}(t)$ crosses $\overline{T}_0$ (when $z(t)$ crosses 1). 

A generalised formulation covering both regimes is obtained by defining a heating rate of the form $v(T,t;R) = \alpha \beta(t; R) (T - R(t))$, where $R \in \{\Tmin, \Tmax \}$ is a reference temperature, which acts as a pivot temperature for the controller, with the property $v(R,t;R)=0$. The reference temperature is defined as follows:
\begin{equation}
R(t)= \begin{cases}
\Tmax, & \text{if } \overline{T}(t) \ge \overline{T}_0 \\
\Tmin, & \text{if } \overline{T}(t) < \overline{T}_0 
\end{cases} 
\end{equation}

\subsection{Control parameter}
The control parameter $\beta(t;R)$ is determined by the desired reference power $\Pi(t)$ according to \cite[Eq.~(36)]{Tindemans2015}:
\begin{align}
\beta(t;R) &=\frac{ \Pi(t)(\Toff - \overline{T}_0)  - (\Toff - \overline{T}(t) ) }{R(t) - \overline{T}(t)} \nonumber \\
&= \frac{(\Pi(t) - 1) - z(t)}{z(t) - \zeta(R(t))}  \label{eq:betaOldDefinition}
\end{align}
where
\begin{equation}
\zeta(R) = \frac{\overline{T}_0 - R}{\Toff - \overline{T}_0}. \label{eq:zetadefinition}
\end{equation}
The denominator in the definition of $\beta$ reflects, in dimensionless form, the energy limits of the TCL aggregate. Note also that $\beta$ switches sign depending on the value of $R(t)$.

\subsection{Distribution scaling}

The controller has the effect of scaling the steady state temperature distribution $f_0(T)$ around the pivot temperature $R(t)$, such that the distribution does not exceed the temperature bounds $\Tmin$ and $\Tmax$ \cite{Tindemans2015}. The extent of this scaling at time $t_i$ is compactly represented by the scale parameter
\begin{align}
s(t) &= \frac{R(t) - \overline{T}(t)}{R(t) - \overline{T}_0} \nonumber \\
&= 1- z(t)/\zeta(R(t)) . \label{eq:sdefinition}
\end{align}

\subsection{Discretisation} \label{sec:discretisation}

We now consider the restriction of the continuous time controller to the set of discrete times $t_i$. We replace the coordinate $z(t)$ by its discretisation $z_i = z(t_i)$, which is computed from \eqref{eq:avgtempdefinition} as 
\begin{equation}
z_i = \sum_{j=0}^{i} (\Pi_j - 1) \left( \e{-\alpha (t_i-t_j)} - \e{-\alpha (t_i - t_{j-1})} \right). \nonumber
\end{equation}
Updates to $z_i$ are efficiently implemented using $z_0=0$ (for a steady state initialisation) and the recursive relation
\begin{equation} \label{eq:zupdate}
z_i = z_{i-1} \e{-\alpha \Delta t_i } + (\Pi_i - 1) (1- \e{-\alpha \Delta t_i }).
\end{equation}

The discretised controller switches modes only at instants $t_i$, so $R(t)$ is approximated by the delayed function
\begin{equation}
\hat{R}(t) = R_i, \quad \text{for } t\in (t_{i-1}, t_i]. \nonumber
\end{equation}
with
\begin{equation} \label{eq:Rip1definition}
R_{i+1} = \begin{cases}
\Tmax, & \text{if } z_i \le 0, \\
\Tmin, & \text{if } z_i >0.
\end{cases}
\end{equation}
Because our analysis focuses on the control time $t_i$, where $\hat{R}(t)$ and $\Pi(t)$ are potentially discontinuous, we introduce $\pm$-notation for the left and right limits at $t_i$: 
\begin{subequations}
\begin{align}
R_i^- &= \lim_{\varepsilon \downarrow 0}  \hat{R}(t-\varepsilon) = R_i, \\
R_i^+ &= \lim_{\varepsilon \downarrow 0}  \hat{R}(t+\varepsilon) = R_{i+1}.
\end{align}
\end{subequations}
Similar definitions using left and right limits naturally apply to $\zeta(R)$, $s(t)$ and $\beta(t;R)$:
\begin{subequations} \label{eq:betasdiscrete}
\begin{align} 
\zeta^{\pm}_{i} & =\frac{\overline{T}_0 - R^{\pm}_{i}}{\Toff - \overline{T}_0}, &
\beta^{-}_{i} &=\frac{(\Pi_i - 1) - z_i}{z_i - \zeta_i^-}, \\
s^{\pm}_{i} &= 1 - z_i /  \zeta_i^{\pm}.&
\beta^{+}_{i} & = \frac{(\Pi_{i+1} - 1) - z_i}{z_i - \zeta_{i}^+}, 
\end{align}
\end{subequations}

\subsection{Energy and power constraints} \label{sec:constraints}

The ability of the aggregate appliances to sustain a low or high power level is determined by operating temperature bounds of the appliance, applied to the distribution-averaged temperature: $\overline{T}(t) \in (\Tmin, \Tmax)$ (no feasible solutions for the distribution $f_z(T)$ exists outside of this domain). However, because operation near the limits is infeasible in practice, due to diverging switching rates, we shall use a restricted range of operating temperatures that is scaled with a fraction $w<1$ around the steady state operating temperature $\overline{T}_0$:
\begin{equation}
(1-w)\overline{T}_0 + w \Tmin \le \overline{T}(t) \le (1-w)\overline{T}_0 + w \Tmax. \nonumber
\end{equation}
Rewriting this in terms of $z(t)$ and $\zeta(\cdot)$, we get
\begin{equation}
w \zeta(\Tmax)  \le z(t) \le w \zeta(\Tmin). \nonumber
\end{equation}
Small excursions out of this temperature band will be permitted, but if this happens, the requested power level $\Pi_{i+1}$ will be restricted to not exacerbate the excursion, using the relation~\eqref{eq:zupdate}. This leads to the update rule for $\Pi_{i+1}$:
\begin{equation}
\Pi_{i+1} := \begin{cases}
\max (\Pi_{i+1}, 1+ w \zeta(\Tmax)), & \text{if } z_i \le w \zeta(\Tmax)\\
\min (\Pi_{i+1}, 1 + w \zeta(\Tmin) ), & \text{if } z_i \ge w \zeta(\Tmin)\\
\Pi_{i+1}, & \text{otherwise} 
\end{cases}
\end{equation}

In addition to energy constraints related to the distribution-averaged temperature, the controller is subject to instantaneous power constraints that result from the maximum rate of change of the distribution. If the controller is in \emph{energy provision mode} ($z_i \le 0$), these power constraints are given by \cite{Trovato2015}
\begin{multline}\label{eq:instantaneousPowerLimits}
 \left(\frac{\overline{T}_0 - \Tmin}{\Tmax-\Tmin} \right) \left( \frac{\Toff - \Tmax}{\Toff - \overline{T}_0} \right) \le \Pi_{i+1} \le \\ \left(\frac{\Toff - \Tmax}{\Toff - \overline{T}_0}\right) + \frac{(\Tmax - \bar{T}_0)(\Tmax - \Ton) }{(\Tmax - \Tmin) (\Toff - \bar{T}_0)}.
\end{multline}
If the controller is in \emph{energy absorption mode} ($z_i > 0$), the power constraints are given by
\begin{multline}\label{eq:instantaneousPowerLimits2}
 \left(\frac{\Tmax - \overline{T}_0 }{\Tmax-\Tmin} \right) \left( \frac{\Toff - \Tmin}{\Toff - \overline{T}_0} \right) \le \Pi_{i+1} \le \\ \left(\frac{\Toff - \Tmin}{\Toff - \overline{T}_0} \right)+ \frac{( \bar{T}_0 - \Tmin)(\Tmin - \Ton) }{(\Tmax - \Tmin) (\Toff - \bar{T}_0)}.
\end{multline}

\section{Device switching} \label{sec:switching}
The desired evolution of the temperature distribution can be used to compute the necessary control actions of individual appliances. This section identifies such control actions using the three types of switching events identified in Section~\ref{sec:drcontrol}. These are computed as a function of the time of evaluation $t_i$, the compressor state $c_i \in \{0,1 \}$ during the preceding interval $(t_{i-1}, t_i]$, and the current device temperature $T_i$ (assumed to be measured in the appliance at time $t_i$). 

\subsection{Forced switching}

TCLs are forced to switch \texttt{on} or \texttt{off} when their temperatures exceed the permitted interval $[\Tlow(t), \Thigh(t)]$. From the linear scaling of the temperature distributions around the pivot temperature $R(t)$ with a factor $s(t)$, it follows that 
\begin{align}
\Tlow(t) = R(t) - (R(t)  - \Tmin)s(t), \nonumber \\
\Thigh(t) = R(t) - (R(t) - \Tmax) s(t). \nonumber
\end{align}
At $t_i$, the refrigerator must act if these bounds are violated at the start of the next time interval:
\begin{align}
\left[T_i \le R_i^+ - (R_i^+  - \Tmin) s^+_i \right], & \Rightarrow c_{i+1} := 0 \\
\left[T_i \ge R_i^+ - (R_i^+  - \Tmax) s^+_i \right]. & \Rightarrow c_{i+1} := 1 
\end{align}

\subsection{Continuous-time switching}

We now consider the continuous-time stochastic switching rates from \texttt{on} to \texttt{off} states ($r^{1\rightarrow 0}(t)$) and vice versa ($r^{0 \rightarrow 1}(t)$), required to maintain the desired shape of the temperature distribution. The switching rates for the energy provision mode are defined in \cite[Eqs.~(48)-(52)]{Tindemans2015}. Here, we generalise these expressions to cover both energy provision and absorption modes ($R(t) \in \{  \Tmin, \Tmax\}$) and simplify them using the $z$-coordinate transformation. Finally, we specialise the expressions for trapezoidal integration with piecewise constant power references.

The derivative of $\beta$ can be simplified by substitution using \eqref{eq:zdefinition} and \eqref{eq:zetadefinition}, resulting in:
\begin{align}
\frac{\mathrm{d}\beta(t;R)}{\mathrm{d} t} &=   \frac{1}{z(t) - \zeta(t)}\frac{\mathrm{d}\Pi(t)}{\mathrm{d}t} + \alpha \beta(t;R) \frac{1 + \zeta(t) - \Pi(t)}{z(t) - \zeta(t)} \nonumber \\
&= \frac{1}{z(t) - \zeta(t)}\frac{\mathrm{d}\Pi(t)}{\mathrm{d}t} - \alpha \beta(t;R) (1+\beta(t;R)) \nonumber
\end{align}
This substitution can be used in \cite[Eq.~(51)]{Tindemans2015} (adjusted for general $R$). Further simplification follows from setting $\mathrm{d}\Pi(t)/ \mathrm{d}t = 0$ (because we consider piecewise constant sections between $t_i$). We compute the intermediate quantity $\Xi(t)$, using the identity found in \cite[Eq.~(38)]{Tindemans2015}, again taking left and right limits due to discontinuity at $t_i$.
\begin{subequations}\label{eq:xidefinition}
\begin{align} 
\Xi^{\pm}_i =& \lim_{\varepsilon \downarrow 0}\Xi(T_i,t_i\pm \varepsilon) \nonumber \\
=& \alpha^2 \left(\frac{P^{\pm}_i + Q^{\pm}_i}{P^{\pm}_i Q^{\pm}_i} \right) (X^{\pm}_i Y^{\pm}_i)  -  \alpha^2 (1+\beta^{\pm}_i) (X^{\pm}_i + Y^{\pm}_i) 
\end{align}
with
\begin{align}
P^{\pm}_i &= (T_i - \Toff) + (\Toff - R^{\pm}_i)(1-s^{\pm}_i)\\
Q^{\pm}_i &= (T_i - \Ton) + (\Ton - R^{\pm}_i)(1-s^{\pm}_i)\\
X^{\pm}_i &= (T_i - \Toff) + (T_i - R^{\pm}_i) \beta^{\pm}_i\\
Y^{\pm}_i &= (T_i-\Ton) + (T_i - R^{\pm}_i) \beta^{\pm}_i 
\end{align}
\end{subequations}
The stochastic transition rates at the left and right limits to $t_i$ are computed from $\Xi^{\pm}_i$ using \cite[Eq.~(49)]{Tindemans2015} (adjusted for general $R$), resulting in
\begin{subequations} \label{eq:rateequations}
\begin{align}
r^{1\rightarrow 0}_{i,\pm} &= \max\left(0,\frac{- \Xi^{\pm}_i}{\alpha(T_i - \Toff) + \alpha \beta^{\pm}_i (T_i - R^{\pm}_i)} \right) \\
r^{0 \rightarrow 1}_{i,\pm} &= \max\left(0,\frac{- \Xi^{\pm}_i}{\alpha(T_i - \Ton) + \alpha \beta^{\pm}_i (T_i - R^{\pm}_i)} \right) 
\end{align}
\end{subequations}

Midpoint integration between adjacent time instants $t_{i-1}$ and $t_i$ is used to determine the resulting switching probabilities, where switching is implemented at $t=t_i$:
\begin{subequations} \label{eq:stochastic}
\begin{align}
\text{Pr}^{1  \rightarrow 0}_{\text{cont},i} &= \frac12 \Delta t_i (r^{1  \rightarrow 0}_{i-1,+} + r^{1 \rightarrow 0}_{i,-})  \\ 
\text{Pr}^{0 \rightarrow 1}_{\text{cont},i} &= \frac12 \Delta t_i (r^{0 \rightarrow 1}_{i-1,+} + r^{0 \rightarrow 1}_{i,-}) 
\end{align}
\end{subequations}
Note that the rates at both `inner' edges of the interval $\Delta t_i$ are used: the `$+$' side at $t_{i-1}$ and the `$-$' side at $t_i$.

\subsection{Instantaneous switching}

Finally, consider the instantaneous stochastic switching at time $t_i$ due to mode changes (energy absorption, energy delivery) or changes in $\Pi(t)$. This results in a discontinuous change in the net heating rate $v(T,t)$, which can only be achieved by a fraction of devices switching on or off at $t_i$. Following \cite{Trovato2015}, we compute the probability of switching from the \texttt{on} to \texttt{off} state at time $t_i$, for a refrigerator that is currently \texttt{on}, as
\begin{subequations} \label{eq:instantaneous}
\begin{equation}
\text{Pr}^{1 \rightarrow 0}_{\text{inst},i}=\text{max}\left(0, 1- \frac{(T_i - \Toff) + (T_i - R_{i}^+)\beta^{+}_i}{(T_i - \Toff ) + (T_i - R_i^-)\beta^{-}_i } \right) \label{eq:instantjumpprob1}
\end{equation}
Note that the switching probability includes both a contribution from the discrete change in power level at $t_i$ as well as a possible mode transition in the previous interval that is implemented at $t_i$. The switching probability for fridges in the \texttt{off} state, $c_i=0$, is defined analogously as
\begin{equation}
\text{Pr}^{0 \rightarrow 1}_{\text{inst},i}=\text{max}\left(0, 1- \frac{(T_i - \Ton) + (T_i - R_{i}^+)\beta^{+}_i }{(T_i - \Ton)+ (T_i - R_i^-)\beta^{-}_i } \right) \label{eq:instantjumpprob2}
\end{equation}
\end{subequations}

\subsection{Combined stochastic switching}
Formally, the continuous-time \eqref{eq:stochastic} and instantaneous \eqref{eq:instantaneous} switching probabilities should be evaluated in sequence, because the former occurs during the interval $(t_{i-1},t_i]$ and the latter at time $t_i$. This would account for the possibility that an appliance switches off and on again within a single interval, or vice versa. Here, we assume that the switching probability associated with the continuous-time process is small to allow us to evaluate both probabilities in a single step. 
\begin{subequations} \label{eq:finalprobabilities}
\begin{align} 
\text{Pr}^{1 \rightarrow 0}_i &=  \text{Pr}^{1  \rightarrow 0}_{\text{cont},i} +  \text{Pr}^{1 \rightarrow 0}_{\text{inst},i} \\
\text{Pr}^{0 \rightarrow 1}_i & = \text{Pr}^{0 \rightarrow 1}_{\text{cont},i} + \text{Pr}^{0 \rightarrow 1}_{\text{inst},i}
\end{align}
\end{subequations}

\section{Algorithm and results}

The discrete time algorithm for updating the compressor state derived in sections~\ref{sec:distribution} and \ref{sec:switching}, is summarised in pseudocode in Algorithm~\ref{alg:update}. The algorithm was implemented in Python 3.7 using the \texttt{numba} package to benefit from just-in-time compilation for considerable speedups.

\begin{algorithm}
\caption{State update algorithm}\label{alg:update}
\begin{algorithmic}[5]
\Function{update\_compressor\_state}{$\Pi_{i+1}, T_i, t_i$}
	\State \emph{\# load previously computed information}
	\State load appliance model $\mathcal{M}$ and operating range $w$
	\State load $c_i, \Pi_i, z_{i-1}, t_{i-1}, r^{1\rightarrow 0}_{i-1,+} , r^{0\rightarrow 1}_{i-1,+} $ 
	\State \emph{\# implement power and energy limits}
  	\State compute $z_i$ using \eqref{eq:zupdate} 
	\If{$z_i \le 0$} \Comment{energy delivery mode}
		\If{$z_i \le w \zeta(\Tmax)$}
		\State $\Pi_{i+1} \gets \max (\Pi_{i+1}, 1+ w \zeta(\Tmax))$
		\EndIf
		\State clip $\Pi_{i+1}$ to limits in \eqref{eq:instantaneousPowerLimits} 
	\Else \Comment{energy absorption mode}
		\If{$z_i \ge w \zeta(\Tmin)$}
			\State $\Pi_{i+1} \gets \min (\Pi_{i+1}, 1 + w \zeta(\Tmin))$
		\EndIf
		\State clip $\Pi_{i+1}$ to limits in \eqref{eq:instantaneousPowerLimits2} 	
	\EndIf
	\State \emph{\# determine distribution and switching variables}
	\State compute $R^{\pm}_{i}, \zeta^{\pm}_{i}, \beta^{\pm}_i, s^{\pm}_i$ using \eqref{eq:Rip1definition}-\eqref{eq:betasdiscrete}
	\State compute $r^{1\rightarrow 0}_{i,\pm}, r^{0\rightarrow 1}_{i,\pm}$  using \eqref{eq:xidefinition}-\eqref{eq:rateequations}
	\State compute $\text{Pr}^{1 \rightarrow 0}_i, \text{Pr}^{0 \rightarrow 1}_i$ using \eqref{eq:stochastic}-\eqref{eq:finalprobabilities}
	\State \emph{\# implement compressor switching}
	\If{$c_i = 1$} \Comment{currently \texttt{on}}
		\If{$T_i \le R_i^+ - (R_i^+ - \Tmin)s^+_i$} $c_{i+1} \gets 0$ 
		\Else
			\State $U \gets \textrm{uniform random}\in [0,1]$
			\If{$U \le \text{Pr}^{1 \rightarrow 0}_i$} $c_{i+1} \gets 0$
			\Else ~$c_{i+1} \gets 1$ \Comment{remain \texttt{on}}
			\EndIf
		\EndIf
	\Else \Comment{currently \texttt{off}}
		\If{$T_i \ge R_i^+ - (R_i^+ - \Tmax)s^+_i$} $c_{i+1} \gets 1$
		\Else
			\State $U \gets \textrm{uniform random}\in [0,1]$
			\If{$U \le \text{Pr}^{0 \rightarrow 1}_i$} $c_{i+1} \gets 1$
			\Else ~$c_{i+1} \gets 0$ \Comment{remain \texttt{off}}
			\EndIf
		\EndIf
	\EndIf
   \State \textbf{return} $c_{i+1}$ \Comment{updated compressor state}
\EndFunction
\end{algorithmic}
\end{algorithm}

For simulations, thermal model parameters were taken from \cite[domestic refrigerator class]{Trovato2015}: $\alpha=1/7200s$; $\Tmax=7\degree C$; $\Tmin=2 \degree C$; $\Ton=-44\degree C$; $\Toff=20 \degree C$. Heterogeneous appliances were generated from these parameters by individually multiplying them with a random factor that was uniformly distributed between $0.8$ and $1.2$. All appliances had a maximum power consumption $P^a_{\mathrm{on}}=70W$ and operating range $w=0.9$ (not binding for the parameters used). Each appliance was randomly initialised as follows. The compressor was set to the \texttt{on} state with a probability equal to the steady state duty cycle $P^a_0/P^a_{\mathrm{on}}$ and the temperature was initialised according to the steady state probability distributions $f_0(T | c^a_0=1) \propto 1/(T-\Ton)$ and $f_0(T | c^a_0=0) \propto 1/(\Toff - T)$. 

\begin{figure}[!t]
\centering
\includegraphics{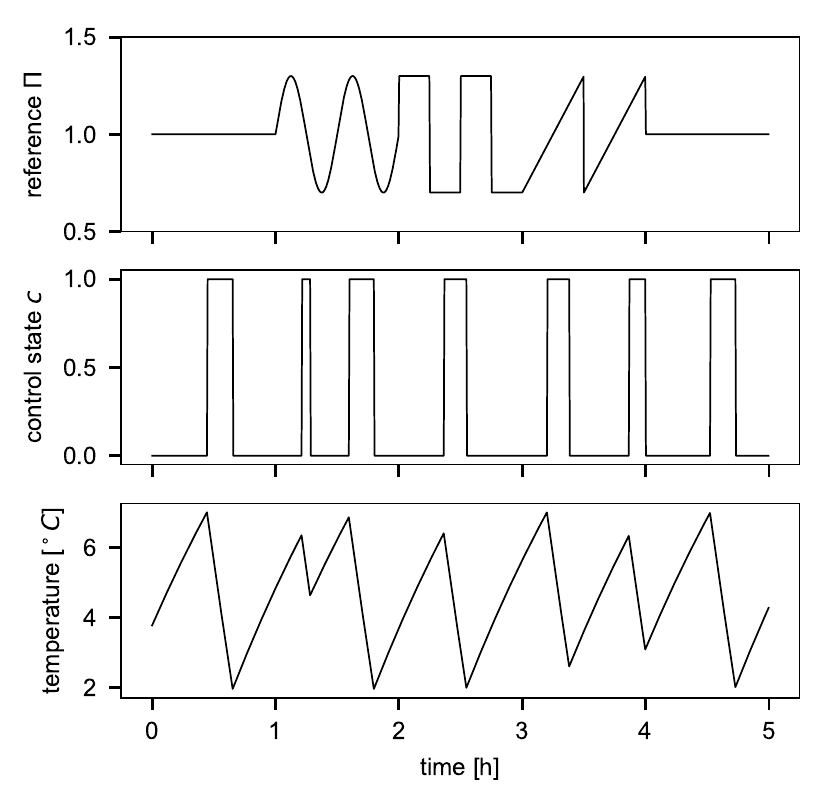}
\caption{Reference signal (top) and response of a single appliance (middle and bottom).}
\label{fig:single}
\end{figure}

Figure~\ref{fig:single} shows a reference signal (top), with a length of 5 hours, that demonstrates a variety of features. The middle and bottom panels show the compressor state $c_i$ and temperature $T_i$, respectively, of a single appliance that tracks the reference signal. These results illustrate the apparently weak relation between the reference signal and single device dynamics (middle), and the ability of the controller to strictly respect the temperature bounds (bottom).

\begin{figure}[!t]
\centering
\includegraphics{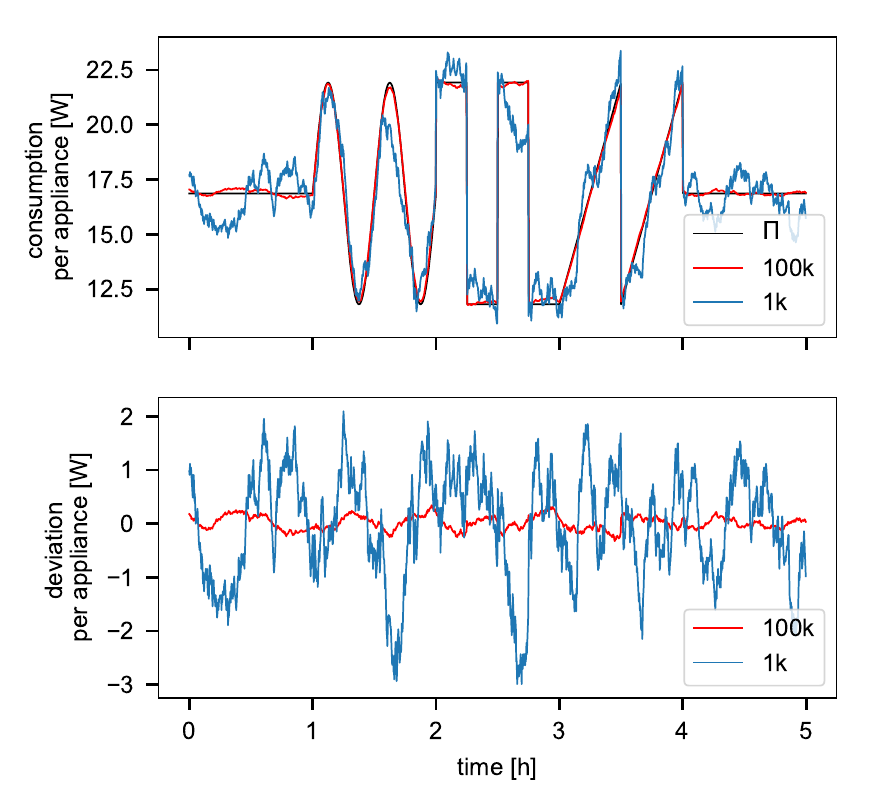}
\caption{Response of a heterogeneous aggregate of appliances (top) and deviation from the reference (bottom).}
\label{fig:multi}
\end{figure}

Next, heterogeneous populations of $1{,}000$ and $100{,}000$ appliances were simulated, tracking the same reference signal. Figure~\ref{fig:multi} illustrates the convergence of the aggregate response to the reference signal as the number of independent appliances increases. The top panel shows absolute power consumption per appliance; the bottom panel the deviation from the reference. 

\section{Conclusions and future work}
This paper has derived a discrete time TCL controller for decentralised demand response. The results illustrate the ability to accurately track a reference signal with a large population of heterogeneous appliances. Moreover, Algorithm~\ref{alg:update}  has low computational complexity, which permits implementation on embedded hardware with severe computational constraints, or it can be used to achieve efficient simulations. The simulation of 100,000 devices for 5 hours using 10s time steps took only 36 seconds (using an Intel i5-7360U CPU under macOS 10.14.2). Moreover, the ability to use variable time steps can further alleviate real time constraints.

Lab testing of the algorithm in a modified refrigerator is currently ongoing. Both in the lab and using simulations, it is of interest to investigate the robustness of the controller against perturbations from the idealised setting \cite{Kara2015}. Initial simulation experiments suggest that the performance of the controller is quite robust to misspecification of the thermal model. Nevertheless, it is interesting to consider how the controller could be enhanced with a means for an appliance to learn and test its own thermal model. 

\bibliographystyle{IEEEtran}
%

\end{document}